\documentclass[12pt,twoside]{article}
\usepackage{fleqn,espcrc1}

\usepackage{graphicx}
\usepackage[figuresright]{rotating}

\def\bra{\,<\!} \def\ket{\!>\,} \def\ack{\,|\,}

\title{Transition quadrupole moments in $\gamma$-soft nuclei
       and the triaxial projected shell model}

\author{Javid A. Sheikh,\address{Physik-Department, Technische
Universit\"at M\"unchen, D-85747 Garching, Germany} 
Yang Sun,\address{Department of Physics and Astronomy, University of
Tennessee\\ Knoxville, Tennessee 37996, U.S.A.}
\address{Department of Physics, Tsinghua University, Beijing
100084, P.R. China}
\address{Department of Physics, Xuzhou Normal University\\
Xuzhou, Jiangsu 221009, P.R. China}
Rudrajyoti Palit\address{Tata Institute of Fundamental
Research, Colaba, Bombay - 400 005, India}
}

\begin{document}
\maketitle

\bigskip
\begin{abstract}
ABSTRACT: Transition probabilities of the ground-state bands
in $\gamma$-soft nuclei are studied for the first time using the
triaxial projected shell model approach. It is observed that 
the angular-momentum dependence of the transition quadrupole moment $Q_t$ 
is related to the triaxial deformation of the nuclear mean-field potential.
The introduction of the $\gamma$-degree of freedom in the
shell model basis is shown to have a little influence 
on the {\it constant} behavior
of the low-spin $Q_t$ in a well-deformed nucleus. However, the {\it increasing} 
collectivity with spin for the low-spin states in a $\gamma$-soft nucleus 
can only be explained by considering the triaxial mean-field deformation.
\end{abstract}

\bigskip
\bigskip

The study of the nuclear shape as a function of angular-momentum 
has remained in the forefront of the nuclear physics research. 
The study of transition probabilities plays an important role in 
our understanding of the shape evolution. For instance, probability of the 
electric quadrupole transition directly reflects the deformation of 
a nuclear system. For a spherical
nucleus, the electric quadrupole transition is of the order of a Weisskopf
unit, whereas for a deformed system the transition is several hundred times 
the Weisskopf estimate \cite{BM75}. 
Nuclear deformation is also believed to be 
one of the most important physical quantities in the
astrophysical interest \cite{Sch98}.  
Many nuclei in nuclear periodic
table exhibit axially symmetric deformation in their ground-state,
with the projection of angular-momentum on the
symmetry-axis as a conserved quantum number. This is referred 
to as the $K$-quantum number and the rotational bands are labelled with
this quantum-number. In fact, the majority of electromagnetic transitions 
in nuclei are found to strictly obey 
the selection rules based on the $K$ quantum-number \cite{BM75}. 
The violation of
$K$-selection rules is an indication that the system is not axially symmetric.
This has been demonstrated, for example, in high-$K$ isomeric states \cite{xwsw98}.

Well-deformed nuclei, in particular those in the heavy mass regions, 
exhibit the characteristics of an axially symmetric-rotor in the low-spin states of their
ground-band (g-band). For example, in well-deformed rare-earth nuclei,
the transition quadrupole moments $Q_t$ usually show a constant behavior
for the spin-range $0\le I\le 10$, before the first band-crossing. 
This behavior has been found in many calculations with the cranked mean-field models 
(see, for example, Ref. \cite{Ces91}) and also in those with angular-momentum projection 
\cite{Sun94,Vel99,Tanabe99}. 

However, the properties of quadrupole moments 
for nuclei in the transitional regions have been less understood. 
There has been a long-standing problem in nuclear structure physics:
what are the finger prints of triaxial deformation in the electromagnetic transition
probabilities in the g-bands for transitional nuclei? 
In the present work, we would like to address this question. It is
demonstrated that the angular-momentum dependence of the transition quadrupole
moments for the low-spin states can provide a measure of the triaxial deformation.

The nuclei discussed here with neutron-number around 90
have g-bands that are quasi-rotational with considerable
vibrational character. The ground-state energy surface of these transitional
nuclei has been shown to have a shallow minimum at a finite
$\gamma$-deformation in the Hartree-Fock-Bogoliubov (HFB)
calculations \cite{KB68}. It was
demonstrated that such a shallow minimum becomes a prominent
one when projected onto spin $I=0$ \cite{HHR82}. 
Thus, to describe these transitional nuclei, it is important to consider the
basis-states which are eigen-states of the triaxial mean-field potential
rather than axial basis used in most of the earlier studies \cite{review}. 
The necessity of introducing triaxiality to describe the observed g-band
moment of inertia of transitional nuclei has recently been demonstrated
\cite{SH99}. The restriction to an axially deformed
basis in the projected shell model approach \cite{review} was released by
performing three-dimensional angular-momentum projection on the
triaxial deformed basis. This approach has been referred to as the triaxial projected
shell model (TPSM). 
It was shown that the observed steep increase of moment of inertia for
transitional nuclei can be well described \cite{SH99} by considering $\gamma$-deformation. 
It was found later \cite{Sun00} that, 
with the same triaxial deformation, the first
excited TPSM band describes also the observed $\gamma$-vibrational band, and
the second excited TPSM band reproduces 
the experimental $\gamma\gamma$-band. 

In the present work, the transition quadrupole moments 
$Q_t$ are studied for the first time by using the 
TPSM. In this approach, the states with 
good angular-momentum are obtained by projection from the triaxial Nilsson
wave-function using the three-dimensional angular-momentum projection
method. Here, our interest lies in the low-spin states with $0\le I\le 10$ 
before the quasi-particle (qp) alignments, and therefore, 
we shall restrict our 
many-body basis to the 
angular-momentum projected triaxial qp-vacuum state:
\begin{equation}
\left\{\hat P^I_{MK}\ack\Phi\ket,~0 \le K \le I \right\},
\label{basis}
\end{equation}
where $\hat P^I_{MK}$ is the projection operator 
\begin{equation}
\hat P^I_{MK} = 
{2I+1 \over 8\pi^2} \int d\Omega\, D^{I}_{MK}(\Omega)\, \hat R(\Omega),
\end{equation}
and $\ack\Phi\ket$ represents the triaxial qp vacuum state. This is
the simplest possible configuration space for an even-even nucleus. It should
be noted that for the case of axial-symmetry, the qp-vacuum state has $K=0$,
whereas in the present case with triaxial deformation, the vacuum state has
all possible $K$-values. The rotational bands based on 
the triaxial vacuum state are obtained by specifying different values
for the $K$-quantum number in the rotational $D$-matrix. 
The allowed values of the $K$-quantum
number for a given intrinsic state are determined through the following
symmetry requirement. For $\hat S = e^{-\imath \pi \hat J_z}$, we have 
\begin{equation}
\hat P^I_{MK}\ack\Phi\ket = \hat P^I_{MK} \hat S^{\dagger} \hat S \ack\Phi\ket
                          = e^{\imath \pi (K-\kappa)} \hat P^I_{MK}\ack\Phi\ket,
\label{symmetry}
\end{equation}
where $ \hat S\ack\Phi\ket = e^{-\imath \pi \kappa}\ack\Phi\ket$. For the 
self-conjugate vacuum state, $\kappa=0$ and, 
therefore, it follows from Eq. (\ref{symmetry})  
that only even-values of $K$ are permitted. 

As in the earlier PSM calculations, we use the pairing plus 
quadrupole-quadrupole 
Hamiltonian \cite{review}
\begin{equation}
\hat H = \hat H_0 - {1 \over 2} \chi \sum_\mu \hat Q^\dagger_\mu
\hat Q^{}_\mu - G_M \hat P^\dagger \hat P - G_Q \sum_\mu \hat
P^\dagger_\mu\hat P^{}_\mu .
\label{hamham}
\end{equation}
The corresponding triaxial Nilsson Hamiltonian 
is given by
\begin{equation}
\hat H_N = \hat H_0 - {2 \over 3}\hbar\omega\left\{\epsilon\hat Q_0
+\epsilon'{{\hat Q_{+2}+\hat Q_{-2}}\over\sqrt{2}}\right\}.
\label{nilsson}
\end{equation}
In Eq. (\ref{hamham}),  
$\hat H_0$ is the spherical single-particle Hamiltonian, which
contains a proper spin--orbit force \cite{NKM}.
For the axial deformation $\epsilon$ in the Nilsson model, 
we take the values given in Ref. \cite{BFM86}.
The interaction strengths are taken as follows: The $QQ$-force
strength $\chi$ is adjusted such that the quadrupole
deformation $\epsilon$ is obtained as a result of the self-consistent
mean-field HFB calculation \cite{review}. The monopole pairing
strength $G_M$ is of the standard form
$G_M = \left[21.24
\mp13.86(N-Z)/A\right]/A$, with ``$-$" for neutrons and ``$+$" for
protons, which approximately reproduces the observed odd--even mass
differences in the mass region. This choice of $G_M$ is appropriate for
the single-particle space employed in the PSM, where three major shells
are used for each type of nucleons ($N=4,5,6$ for neutrons and $N=3,4,5$
for protons, a model space appropriate for normally deformed rare earth nuclei).
The quadrupole pairing strength $G_Q$ is assumed to be
proportional to $G_M$, the proportionality constant being fixed as usual
to be in the range 0.16 -- 0.18. These interaction strengths are
consistent with those used previously for the same mass region
\cite{review,SH99,Sun00}.

The Hamiltonian in Eq. (\ref{hamham}) is diagonalized using the projected basis
of Eq. (\ref{basis}). The obtained wave-function can be written as
\begin{equation}
\Psi^\sigma_{IM} = 
\sum _{K} f^{\sigma}_{IK}\,\hat P^I_{MK}\ack\Phi\ket .
\end{equation} 
Note that, although only the qp-vacuum state is included in the basis
in Eq. (\ref{basis}), its triaxial nature generates the 
$K$-state mixing when the diagonalization is carried out. 
The expansion coefficients $f$, obtained through the 
diagonalization of the shell-model Hamiltonian, 
describe the amount of $K$-mixing
and specify various physical states (e.g. g-, $\gamma$-, $\gamma\gamma$-bands)
\cite{Sun00}.
The wave-function is then used to evaluate the electromagnetic 
transition probabilities. 
The reduced electric transition probabilities $B(EL)$ from an initial state 
$( \sigma_i , I_i) $ to a final state $(\sigma_f, I_f)$ are given by 
\begin{equation}
B(EL,I_i \rightarrow I_f) = {\frac {1} {2 I_i + 1}} 
| \bra \Psi^{\sigma_f}_{I_f} || \hat Q_L || \Psi^{\sigma_i}_{I_i} \ket |^2 ,
\end{equation}
and the reduced matrix element can be expressed as  
\begin{eqnarray*}
&\bra& \Psi^{\sigma_f}_{I_f} || \hat Q_L || \Psi^{\sigma_i}_{I_i} \ket
\nonumber \\ 
&=& \sum_{K_i,K_f} f_{I_i K_i}^{\sigma_i} f_{I_f K_f}^{\sigma_f}
\sum_{M_i , M_f , M} (-)^{I_f - M_f}
\left(
\begin{array}{ccc}
I_f & L & I_i \\
-M_f & M &M_i 
\end{array} \right) 
\nonumber \\
 & & \times \bra \Phi | {\hat{P}^{I_f}}_{K_f M_f} \hat Q_{LM}
\hat{P}^{I_i}_{K_i M_i} | \Phi \ket 
\nonumber \\
 &=& 2\sum_{K_i,K_f} f_{I_i K_i}^{\sigma_i} f_{I_f K_f}^{\sigma_f}
\nonumber \\
 & & \times \sum_{M^\prime,M^{\prime\prime}} (-)^{I_f-K_f} (2 I_f + 1)^{-1}
\left( 
\begin{array}{ccc}
I_f & L & I_i \\
-K_{f} & M^\prime & M^{\prime\prime}
\end{array} \right) 
\nonumber \\
 & & \times \int d\Omega {\it D}_{M'' K_{i}} (\Omega) 
\bra \Phi | \hat Q_{LM'} \hat{R}(\Omega) | \Phi \ket . 
\end{eqnarray*}
The transition quadrupole moment $ Q_t (I) $ is related to $B(E2)$ transition 
probability through  
\begin{equation}
Q_t(I) = \sqrt{\frac{16 \pi}{5}} \frac{ \sqrt{B(E2,I\rightarrow I-2)}} 
{\bra I, 0, 2, 0 | I-2, 0\ket } . 
\end{equation}
In the calculation, 
we have used the standard effective charges of 1.5e for protons 
and 0.5e for neutrons \cite{Sun94,review}.  

The variation of $Q_t$ as a function of spin $I$ provides the information
about the shape evolution of a rotating nucleus.
In the simplest approximation for a nucleus as 
an axially deformed rigid-body, 
$Q_t$ has a constant value for all the spin-states in a given band. In fact, 
for well deformed nuclei, one finds more or less a constant value
of $Q_t$ at low-spin states up 
to the first band-crossing region. Experimentally,
deviations are observed from the rigid-body behavior 
in many nuclei at high-spin states, especially in  
the band-crossing region -- one often observes  
a drop in $Q_t$ due to small overlap
between the wave-functions of the initial and the final states involved. 
It was shown \cite{Ces91} 
that the drop in $Q_t$ in the band-crossing region cannot be
quantitatively described by various types of cranking models  
where angular momentum is not treated as a good quantum-number, 
but the calculations based on the projected shell model 
with axially symmetric basis can 
reproduce this phenomenon \cite{Sun94,Vel99}.  
Very recently, it has been reported that angular momentum projection
on the cranked HFB states can also describe the drop \cite{Tanabe99}. 
In the present work, we shall discuss the spin states below the band-crossing region.

The constant behavior of 
$Q_t$ till the band-crossing region is noticed for most of the nuclei
that are well deformed in the ground-state. In comparison,
for some light rare-earth nuclei that belong to the transitional region, 
for instance, $^{154-158}$Dy and $^{156-160}$Er, 
$Q_t$ depicts a variation with angular-momentum for the low-spin
states of the g-band, well below the band crossing region.  
The existing experimental data seem to indicate a general 
trend of increasing $Q_t$ for the spin region $0\le I \le 10$. 
This variation of $Q_t$  cannot be explained by calculations 
using the projected shell model with axially-symmetric basis \cite{Sun94,Vel99}.

\begin{figure}[htb]
\begin{minipage}[t]{140mm}
\includegraphics[scale=0.45]{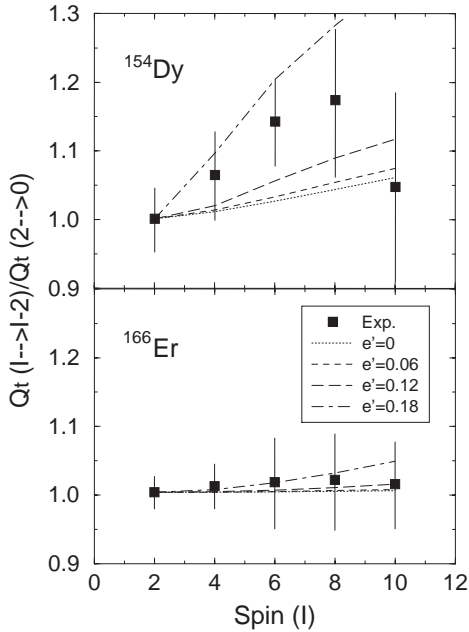}
\caption{Calculated transition quadrupole moments for the transitional nucleus
$^{154}$Dy (top panel), and well-deformed nucleus $^{166}$Er
(bottom panel) with different
triaxial deformation $\epsilon'$.
The corresponding experimental data \protect\cite{Dy154,Er166} are
also shown for a comparison.
}
\label{fig:largenenough}
\end{minipage}
\end{figure}

In Fig. 1, the results of $Q_t$ are presented for various values of the
triaxial deformation parameter $\epsilon'$ 
to investigate the dependence
of $Q_t$ on the triaxiality of the deformed Nilsson basis. 
$\epsilon' =0$ in Fig. 1 corresponds to the 
axially symmetric case. 
The calculated $Q_t(I\rightarrow I-2)$ are normalized to 
the lowest transition $Q_t(2\rightarrow 0)$ in order to demonstrate more clearly
any changes in $Q_t$ as a function of spin.
$^{154}$Dy has neutron number 88 and represents a transitional nucleus.
Experimentally, a steady increase of $Q_t$ has been observed for the several 
low-spin states \cite{Dy154} 
(the drop of $Q_t$ at $I=10$ is due to the first
band-crossing). As can be seen in the top panel of Fig. 1, 
this feature cannot be
obtained when an axially symmetric basis ($\epsilon' =0$) 
is employed in the calculation.
With increasing $\epsilon'$ values, the calculated $Q_t$ curves become steeper,
and eventually the observed $Q_t$ can be reproduced with $\epsilon' \approx 0.15$.

In the bottom panel of Fig. 1, 
similar calculations have been done for a well deformed nucleus
$^{166}$Er having neutron number 98. The measured $Q_t$
shows a rather constant behavior for the low-spin region \cite{Er166}. 
It is interesting to observe very different calculated curves 
from those in the top panel of Fig. 1:
with increasing $\epsilon'$ values, the calculated $Q_t$ curves keep
showing a nearly constant behavior. Slightly enhanced values are obtained only
with the largest $\epsilon'$ in the calculation. We can thus conclude that
the triaxial basis has no significant effect on the g-band properties
for a well deformed nucleus that exhibits an axial rotor behavior
near the ground-state.    

\begin{figure}[htb]
\begin{minipage}[t]{80mm}
\includegraphics[scale=0.45]{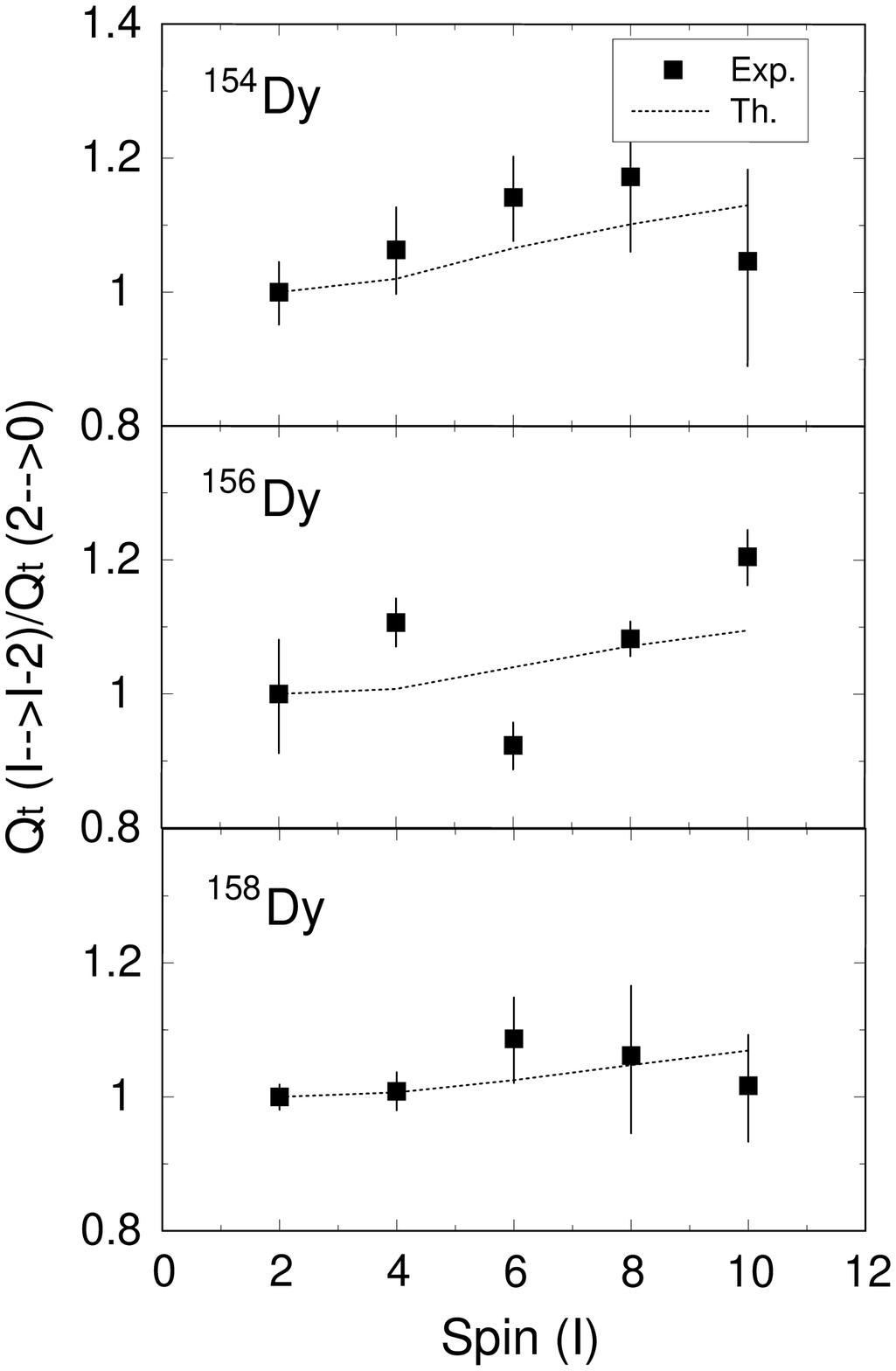}
\caption{Comparison of calculated transition quadrupole moments for
$^{154,156,158}$Dy
(dotted curves)
with the available experimental data \protect\cite{Dy154,Dy156}
(filled squares).}
\label{fig:largenenough}
\end{minipage}
\hspace{\fill}
\begin{minipage}[t]{75mm}
\includegraphics[scale=0.45]{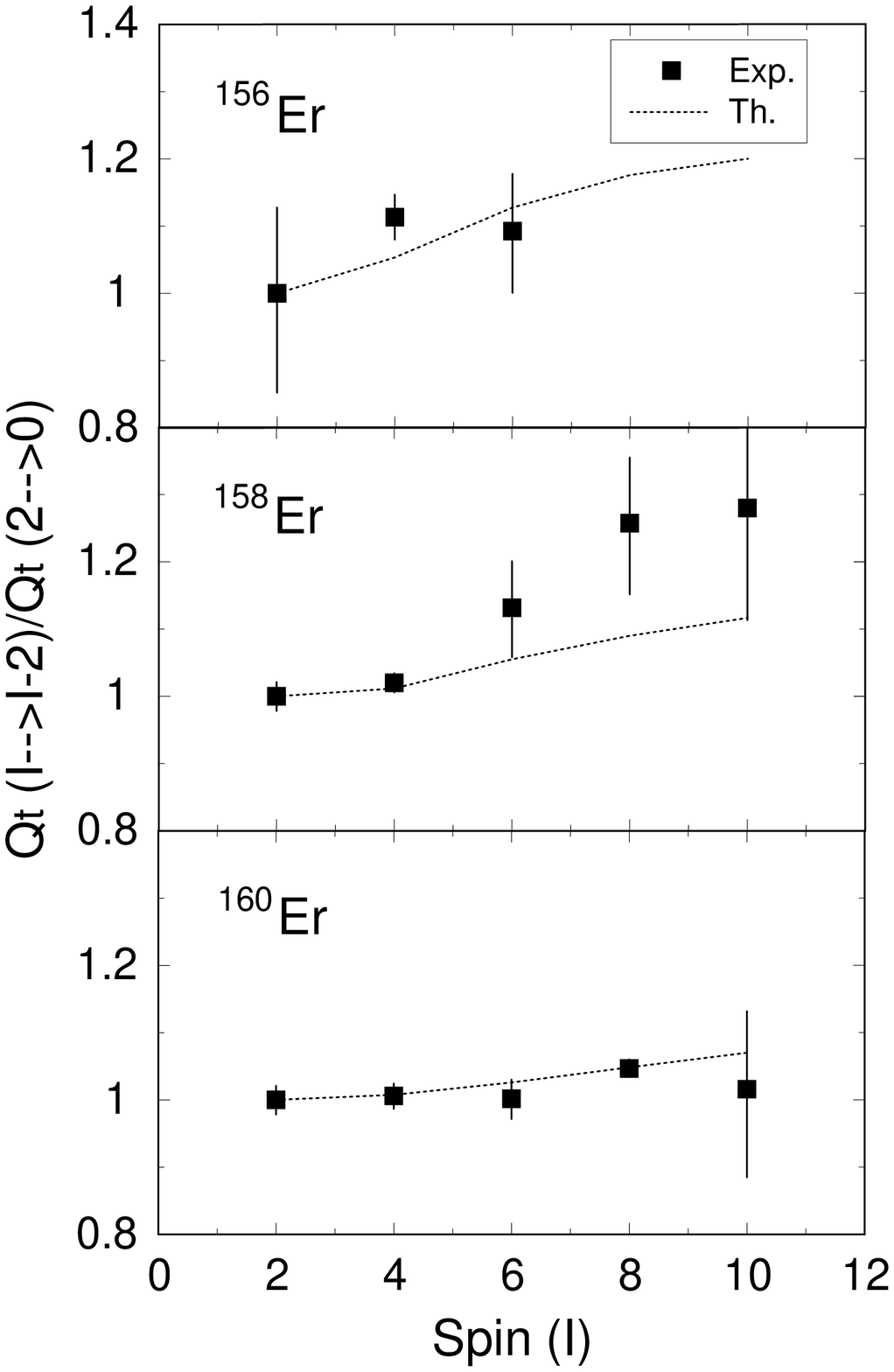}
\caption{Comparison of calculated transition quadrupole moments for
$^{156,158,160}$Er
(dotted curves)
with the available experimental data \protect\cite{Er156,Er158,Er160}
(filled squares).}
\label{fig:toosmall}
\end{minipage}
\end{figure}

In Figs. 2 and 3, the results of the TPSM calculations 
are compared with the available experimental
data for Dy- and Er-isotopes. In order to demonstrate the feature of 
increasing $Q_t$, we present
them in the normalized form $Q_t(I\rightarrow I-2) / Q_t(2\rightarrow 0)$ as in Fig. 1. 
(We would like to add that our calculations also reproduce the absolute values of
$Q_t$ for all the nuclei studied quite accurately).
It is found that by employing a triaxial deformation $\epsilon'\approx 0.15$,
the observed feature of the $Q_t$'s can be reproduced reasonably well in most of
the cases. At this stage, a fine adjustment of $\epsilon'$ in the calculations
for each individual nucleus 
is not necessary since in many of the data discussed here, 
the experimental uncertainties are quite large. 

As is clear in Fig. 2. the experimental $Q_t$ for 
$^{154}$Dy \cite{Dy154} can be nicely reproduced till $I=8$. This was not possible
in the projected shell model calculations if axially deformed basis
was used \cite{Sun94,Vel99}. 
The general feature of the observed $Q_t$ for $^{156}$Dy \cite{Dy156}
also shows an increasing trend with increasing spin, which is described
by the present calculations. The physical reasons for a sudden drop
in the experimental $Q_t$ at $I=6$ are unclear.    
In $^{158}$Dy, the increase in the experimental $Q_t$ 
\cite{Dy156} is not as significant 
as in the two lighter isotopes. This feature has also been reproduced
in our calculations with the same $\epsilon'$ used 
for the $^{154}$Dy and $^{156}$Dy calculations. 

The results for three light Er-isotopes are presented in Fig. 3.
The experimental $Q_t$ of the first three spin-states 
in $^{156}$Er \cite{Er156} 
exhibit an increasing trend in $Q_t$ as a function of spin, which 
is correctly described. The $Q_t$  
for $^{158}$Er were measured in Ref. \cite{Er158} 
up to very high-spin states. The data clearly depict an 
increasing trend, which is reasonably described by the calculation. 
The data for $^{160}$Er \cite{Er160} depict 
nearly constant values for the three low-spin states, with slight variations
for $I=8$ and 10. The agreement of our calculation for this nucleus is
also satisfactory.  
We remark that with the configuration space and interaction strengths
employed in the present work,  
the calculated moments of inertia for the g-bands and the 
calculated spectra for the multi-phonon $\gamma$-vibrational bands were shown to
agree well with the 
observed data \cite{SH99,Sun00}. 

The behavior of $Q_t$
as a function of angular momentum may provide a measure of the
triaxiality for a nuclear system. 
For the transitional nuclei studied in the present
work, larger increase in $Q_t$ with angular momentum
corresponds to higher values of triaxiality used in the basis. The triaxial
deformation parameter $\epsilon'$ is approximately related to the
conventional triaxial parameter $\gamma$ through the relation
$\tan \gamma = {\epsilon \over \epsilon'}$. The value of $\epsilon$ is
held fixed for each nucleus in the calculations, 
and therefore, $\gamma$ increases linearly
with $\epsilon'$. For those well deformed nuclei, the dependence of $Q_t$ on 
the basis triaxiality is insensitive. Therefore, the g-band $Q_t$ alone
are not sufficient to determine triaxiality for a well deformed nucleus,  
and additional physical quantities must be
supplied \cite{Er166,Ring82}.

In conclusion, 
the low-spin transition quadrupole moments of the ground-state bands  
in $\gamma$-soft nuclei have been studied using the
triaxial projected shell model approach. The shell model diagonalization is
carried out with three-dimensional angular-momentum projection based on the
triaxial deformed Nilsson-states. It has been shown that for transitional nuclei, 
the rotational 
evolution of the transition quadrupole moments depends sensitively on  
triaxial deformation of the mean-field potential.

The development of the three-dimensional angular-momentum projection method
for the electromagnetic transition calculations 
has opened possibility of the application of the 
projected shell model approach to a wide range of problems. The next important
application would be the study of the inter-band transitions between
the ground-state band and the $\gamma$-bands. This work is presently 
being pursued and the results will be published in the near future.

\baselineskip = 14pt
\bibliographystyle{unsrt}

\end{document}